# Interfacial fatigue fracture of pressure sensitive adhesives

Yichen Wan[a], Qianfeng Yin[a], Ping Zhang[b], Canhui Yang[b], Ruobing Bai[a,*]

[a] Department of Mechanical and Industrial Engineering, Northeastern University, Boston, MA 02115, USA

[b] Shenzhen Key Laboratory of Soft Mechanics & Smart Manufacturing, Department of Mechanics and Aerospace Engineering, Southern University of Science and Technology, Shenzhen 518055, PR China

[*]Corresponding author: ru.bai@northeastern.edu

**Abstract:** Pressure sensitive adhesives (PSAs) are viscoelastic polymers that can form fast and robust adhesion with various adherends under fingertip pressure. The rapidly expanding application domain of PSAs, such as healthcare, wearable electronics, and flexible displays, requires PSAs to sustain prolonged loads throughout their lifetime, calling for fundamental studies on their fatigue behaviors. However, fatigue of PSAs has remained poorly investigated. Here we study interfacial fatigue fracture of PSAs, focusing on the cyclic interfacial crack propagation due to the gradual rupture of noncovalent bonds between a PSA and an adherend. We fabricate a model PSA made of a hysteresis-free poly(butyl acrylate) bulk elastomer dip-coated with a viscoelastic poly(butyl acrylate-co-isobornyl acrylate) sticky surface, both crosslinked by poly(ethylene glycol) diacrylate. We adhere the fabricated PSA to a polyester strip to form a bilayer. The bilayer is covered by another polyester film as an inextensible backing layer. Using cyclic and monotonic peeling tests, we characterize the interfacial fatigue and fracture behaviors of the bilayer. From the experimental data, we obtain the interfacial fatigue threshold (4.6 J/m$^2$) under cyclic peeling, the slow crack threshold (33.9 J/m$^2$) under monotonic peeling, and the adhesion toughness ($\sim$ 400 J/m$^2$) at a finite peeling speed. We develop a modified Lake-Thomas model to describe the interfacial fatigue threshold due to noncovalent bond breaking. The theoretical prediction (2.6 J/m$^2$) agrees well with the experimental measurement (4.6 J/m$^2$). Finally, we discuss possible additional dissipation mechanisms involved in the larger slow crack threshold and much larger adhesion toughness. It is hoped that this study will provide new fundamental knowledge for fracture mechanics of PSAs, as well as guidance for future tough and durable PSAs.

# 1. Introduction

*Pressure sensitive adhesives* (PSAs) are made of viscoelastic polymer networks in their rubbery phase [1-4]. They form strong adhesion to various materials within seconds under fingertip pressure, without additional chemical or physical treatments. Their softness (typical modulus $\leq 0.1$ MPa) ensures conformal contact to form dense noncovalent bonds at the interface during bonding [5]. During debonding, their viscoelasticity dissipates energy and prevents liquid-like flow under static load.

Over the years, PSAs have been widely used as load bearing tapes in various applications because of their easy handling and fast establishment of adhesion. New PSAs are further being rapidly developed for emerging applications, in fields such as aerospace [6], healthcare [7-10], wearable electronics [11-13], and flexible displays [14-16]. The expanding application domain of PSAs has raised a critical need for their enhanced *fatigue* resistance. Early applications of PSAs, such as duct tapes and Post-it notes, involve mostly static loads, and have focused on their abilities to easily attach or detach. By contrast, new applications require PSAs to maintain their adhesion under prolonged cyclic loads throughout lifetime. Emerging examples include wearable sensors that undergo continuous motions of human bodies, and flexible displays that are required to sustain cyclic folding and unfolding without delamination of functional layers.

Fatigue occurs universally for all load bearing materials and may refer to various symptoms that hinder the material longevity under prolonged *static* or *cyclic* loads [17]. Specifically, under prolonged cyclic loads, some PSAs suffer *fatigue damage*, with their material properties degrading over cycles due to accumulated internal bond breaking. Some other PSAs suffer *bulk fatigue fracture* or *interfacial fatigue fracture*, where cracks nucleate and gradually propagate within the bulk PSA or along the adhesion interface, respectively. These rich phenomena of fatigue in PSAs under cyclic loads, however, have remained poorly investigated, contrasting with the active research on the fracture and debonding of PSAs under static or monotonic loads [18-21].

This paper focuses on interfacial fatigue fracture of PSAs under prolonged cyclic loads, a consequence of the gradual rupture of *noncovalent bonds* between a PSA and an adherend. For *covalently bonded* soft materials and adhesion systems, their bulk and interfacial fatigue fracture have been extensively observed and investigated for decades [3, 17, 22]. Detailed studies have been conducted on traditional rubbers [23-25], single-network hydrogels [26, 27], double-network hydrogels [28-30], double-network elastomers [31], and covalently bonded adhesion interfaces between two soft materials [32, 33]. In these studies, the bulk or interfacial *fatigue threshold* is identified as the lowest energy release rate required to propagate a crack over a reasonably achievable experimental time window. The fatigue threshold of a covalently bonded network or interface is shown to be on the order of 10-50 $J/m^2$ for most soft materials, much lower than their fracture toughness measured under monotonic load. This fatigue threshold and its covalent nature has been further described by the widely known Lake-Thomas model [23]. The extensive research on fatigue fracture has enabled covalently bonded soft materials and adhesion systems with greatly enhanced fatigue thresholds as large as over $10^3$ $J/m^2$ in recent years [34-37]. By contrast, interfacial fatigue fracture of PSAs, along with their interfacial fatigue threshold due to noncovalent bond breaking, has never been investigated.

This paper studies interfacial fatigue fracture of PSAs. We synthesize a model PSA made of a hysteresis-free poly(butyl acrylate) bulk elastomer dip-coated with a viscoelastic poly(butyl acrylate-co-isobornyl acrylate) sticky surface, both crosslinked by poly(ethylene glycol) diacrylate. We adhere the fabricated PSA to a polyester strip to form a bilayer. The bilayer is covered by another polyester film as an inextensible backing layer. We conduct cyclic peeling tests on the bilayer and measure the interfacial crack growth over cycles. We plot the steady-state crack growth per cycle against the applied energy release rate and identify the interfacial fatigue threshold. In addition, we conduct monotonic peeling tests on the same bilayer at different crack speeds, plot the crack speed against the applied energy release rate, and identify the slow crack threshold under monotonic peeling. The interfacial fatigue threshold under cyclic peeling is measured to be 4.6 $J/m^2$, significantly lower than the slow crack threshold of 33.9 $J/m^2$, and further below the adhesion toughness of about 400 $J/m^2$ at finite crack speeds. To understand the measured interfacial fatigue threshold, we develop a modified Lake-Thomas model to describe the interfacial fatigue threshold due to noncovalent bond breaking. The theoretical prediction (2.6 $J/m^2$) agrees well with the experimental measurement (4.6 $J/m^2$). Finally, we discuss possible additional dissipation mechanisms involved in the larger slow crack threshold and much larger adhesion toughness.

## 2. Experimental section

Butyl acrylate (BA, monomer) was purchased from Fisher Scientific. Isobornyl acrylate (IBA, monomer), poly(ethylene glycol) diacrylate (PEGDA, Mn575, crosslinker), 2-Hydroxy-2-methylpropiophenone (I-1173, initiator), ethanol (solvent), and benzophenone (BP, initiator) were purchased from Sigma Aldrich. Polyester film (polyethylene terephthalate, PET, 0.001 inch) was purchased from McMaster-Carr. Silicon oil coated polyester *release film* was purchased from Alibaba. The release film prevents the undesired adhesion between the mold and the prepared sample after its UV curing. All chemicals were directly used as received without further purification.

We fabricate a model PSA following a recent work [38]. The PSA consists of a nearly hysteresis-free elastomeric bulk, dip-coated with a viscoelastic sticky surface layer. The fabrication process is illustrated in **Figure 1**.

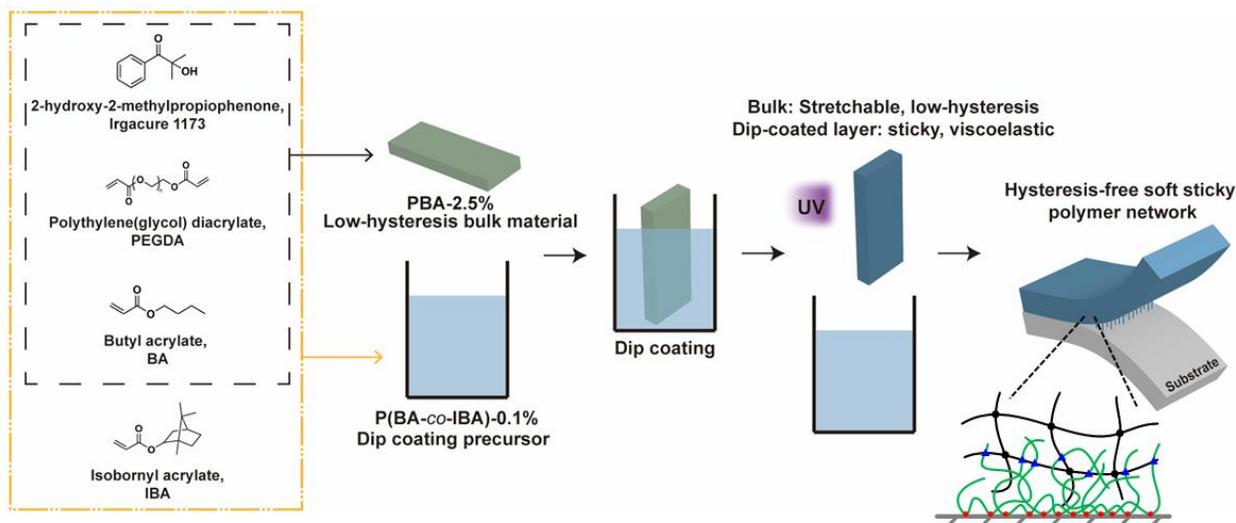

Figure 1. Schematics illustrate the fabrication process of the model PSA. The PSA consists of a nearly hysteresis-free elastomeric bulk, dip-coated with a viscoelastic sticky surface layer.

**2.1 Synthesis of bulk elastomer**

The bulk material of our model PSA is made of a poly(butyl acrylate) (PBA) elastomer. The weight ratios of BA, PEGDA, and I-1173 are prescribed as 100:2.5:1. The three chemicals are mixed by a Vortex mixer for 30 seconds at room temperature to form a homogeneous solution. The solution is subsequently injected into a 10×10 cm$^2$ release film-coated glass mold, separated by a silicone spacer of 1 mm thickness. Afterwards, the solution is cured under the UV light (365 nm) at ambient environment for 30 min. The sample is then taken out and cut into films of 1×6 cm$^2$ using a laser cutter (Epilog 80W). All samples are stored in a sealed plastic box before further dip coating.

**2.2 Dip coating of surface layer**

We prepare a dip coating precursor by first mixing BA, IBA, PEGDA, and I-1173 of 32 g, 8 g, 0.04 g, and 0.04 g, respectively, forming weight ratios of 80:20:0.1:0.1. These chemicals are mixed following the same method in Section 2.1. The mixture is then cured under the UV light for 5 min in a nitrogen atmosphere to form a viscous liquid for dip coating.

In the dip coating, the bulk PBA is first immersed in a benzophenone solution (2 wt% in ethanol) for 1 min. The sample is subsequently taken out and dried in an oven at a fixed temperature of 65 °C for 1 min. During this process, benzophenone initiates free radicals from the C-H bonds on the PBA polymer chains. Afterwards, the bulk PBA is immersed in the prepared dip coating precursor for 1.5 min, taken out, and dried in air for 2 min. The sample is then covered by a PET film on one side, a release film on the other side, and exposed to the UV light for 30 min from the side of the release film. The PET film is left to remain on the fabricated PSA, serving as the inextensible backing layer for the peeling test. The release film can be easily peeled off after curing without contaminating the surface, such that the exposed surface is used to form adhesion with an adherend. The thickness of the sample before and after dip coating is measured by a caliper, showing the coated layer around 250 μm thick on each side.

### 2.3 Preparation of bilayer

To prepare the adhesion bilayer for a peeling test, a PSA strip of size 6 cm ×1 cm ×1.5 mm is adhered to a PET strip of the same size under gentle pressure. The PSA is covered by the other PET film on its back from dip coating, serving as the inextensible backing layer. We form a precut interfacial crack of 0.5 cm by placing a spacer between the PSA and the PET adherend prior to their attachment. The bilayer is subsequently compressed by a 250 g weight for 12 hours in ambient environment before the peeling test.

### 2.4 Uniaxial tensile test

For the uniaxial tensile test, the bulk PBA elastomer without dip coating is cut into a dog bone shape with a gauge length of 50 mm, width of 3 mm, and thickness of 1 mm. The sample is mounted to an Instron tensile tester (Instron 34TM-5, load cell: 100 N), and subjected to cyclic uniaxial tensile tests, with a stretch between 1 and 1.8, at a fixed stretch rate of 1 min$^{-1}$ for 50 cycles.

## 2.5 Cyclic peeling

We conduct cyclic peeling tests on the bilayer sample using the 180-degree setup (**Figure 2a**) [37, 39]. In each peeling test, the tensile tester pulls the two arms of bilayer cyclically with the peeling force between 0 and a prescribed peak force $F$ (**Figure 2b**), under a fixed peeling speed of 30 mm/min. The inextensible backing layer constrains the deformation in the bulk peeling arms, such that the amplitude of the energy release rate, $G$, is calculated as $G = 2F/w$, where $w = 1$ cm is the width of the bilayer.

In all tests, we observe that the crack only propagates along the adhesion interface, without deflecting into the bulk adhesive or adherend, leaving no residual polymers on the adherend after the crack propagation. **Movies 1&2** illustrate peeling tests under cyclic and monotonic loads, respectively. Because of the inextensible backing layer, the crack growth in each cycle can be directly measured as half of the peeling displacement measured by the tensile tester. This measured crack growth agrees well with the displacement calculated from the artificial marks made on the adherend, as shown in **Movies 1&2**. After thousands of cycles, the cyclic peeling reaches a steady state, where the crack growth per cycle is approximately constant. We extract this steady-state crack growth per cycle, $dc/dN$ (**Figure 2c**), and plot it as a function of the prescribed energy release rate $G$ (**Figure 2d**). When $G$ is very small, $dc/dN$ is around 0.5 μm/cycle, and is assumed to vary linearly with $G$. Using linear regression of the data points, the fitted curve intercepts the $G$ axis at a finite value. This value, denoted as $\Gamma_0$, is the measured *interfacial fatigue threshold*.

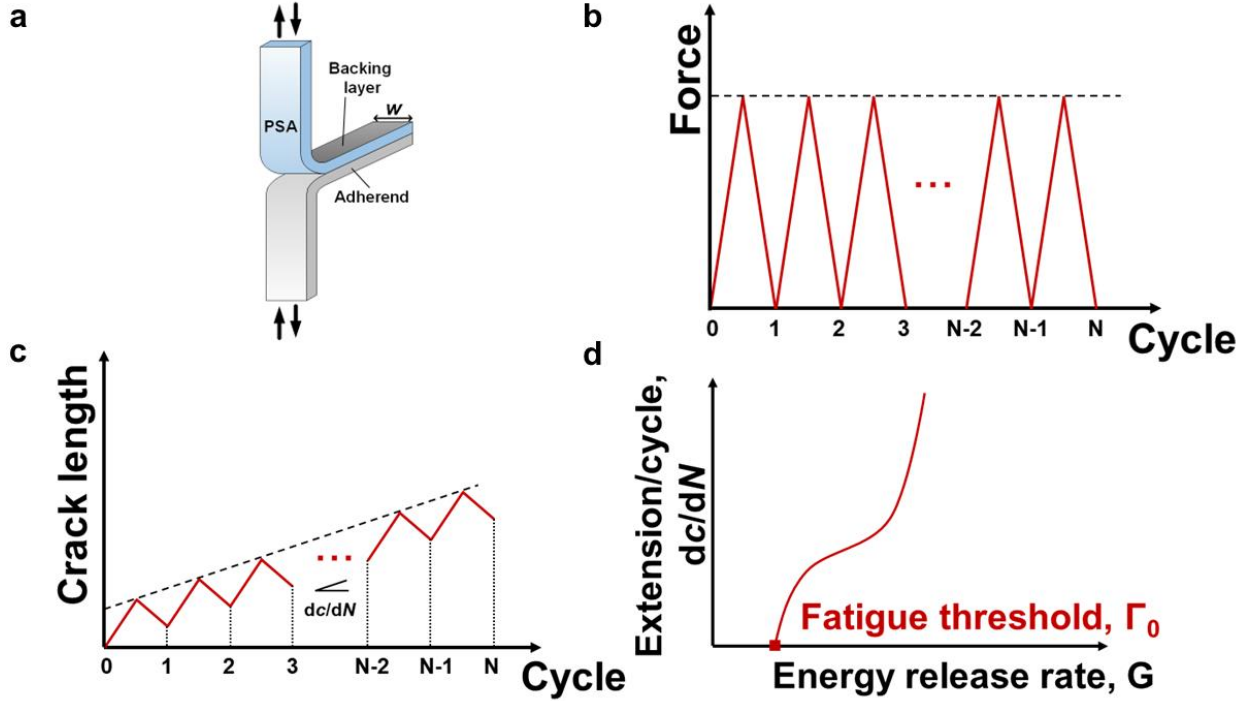

Figure 2. Cyclic peeling tests. Schematics illustrate: (a) the 180-degree peeling setup, (b) the prescribed force-cycle curve, (c) the measured displacement-cycle curve and the steady-state crack growth per cycle, d$c$/d$N$, and (d) d$c$/d$N$ plotted as a function of the prescribed energy release rate $G$, where the interfacial fatigue threshold $\Gamma_0$ is obtained by linear extrapolation.

**2.6 Monotonic peeling**

We conduct monotonic peeling tests using the same 180-degree setup (**Figure 3a**) [22, 40]. In each test, the tensile tester prescribes a constant crack speed $v$ as half of the peeling speed due to the inextensible backing layer (**Figure 3b**). The measured force-displacement curve reaches a steady-state plateau, where the peeling force $F$ is approximately constant (**Figure 3c**). The steady-state energy release rate $G$ is calculated using the same equation $G = 2F/w$. The crack speed $v$ is plotted as a function of $G$ (**Figure 3d**). As $v$ approaches zero, we use the similar linear extrapolation to obtain the finite threshold $G_0$ for monotonic peeling and denote it as the *slow crack threshold* [41].

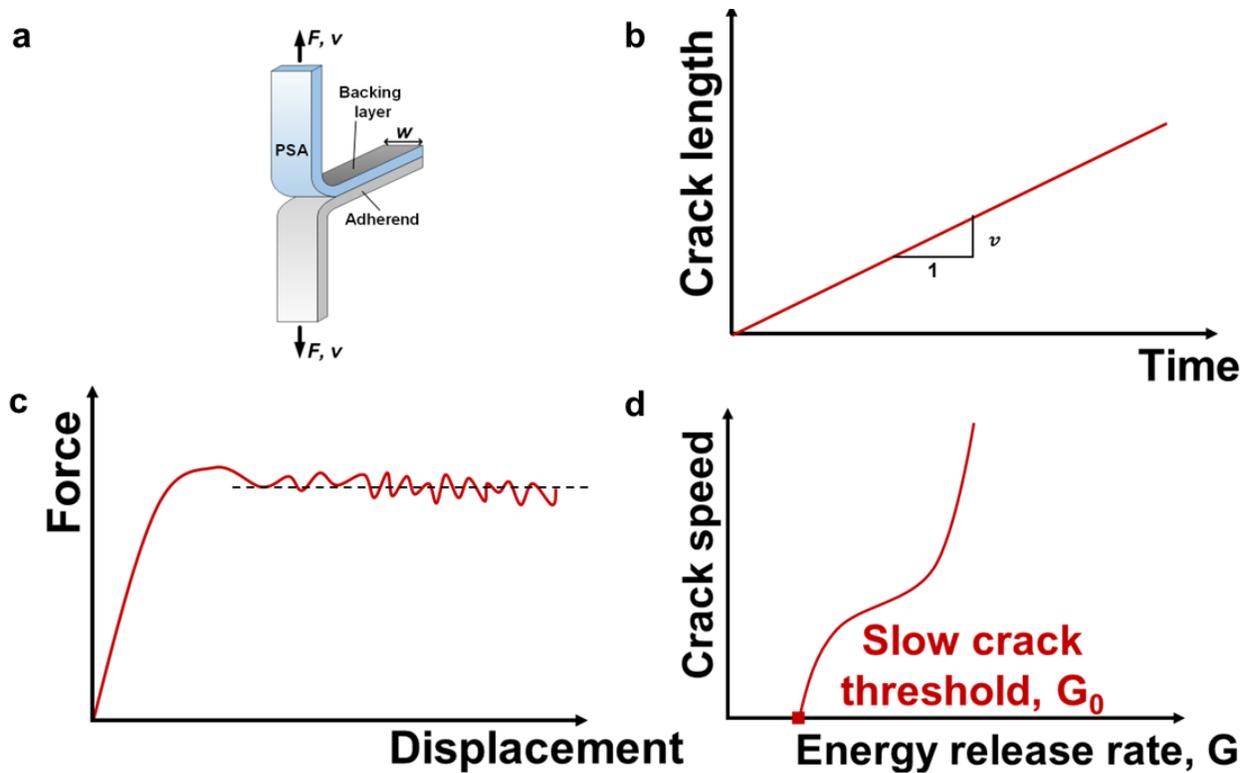

Figure 3. Monotonic peeling tests. Schematics illustrate: (a) the 180-degree peeling setup, (b) the prescribed constant crack speed $v$ as half of the peeling speed due to the inextensible backing layer, (c) the measured force-displacement curve and the steady-state plateau, and (d) $v$ plotted as a function of the steady-state energy release rate $G$, where the slow crack threshold $G_0$ is obtained by linear extrapolation.

For both thresholds measured under cyclic and monotonic peeling, we use a linear extrapolation method. Alternatively, a threshold may be directly represented by the lowest measurable energy release rate in an experiment. We note that this latter approach has also been commonly taken in both cyclic [42] and monotonic [43] tests.

## 3. Mechanical characterizations of bulk elastomer and surface layer

Just like toughening mechanisms in various other soft elastomers and gels [22], the possible high adhesion toughness of a PSA results from energy dissipation at multiple length scales. During debonding, densely packed physical bonds bridge the crack, transmit the local high stress to a large volume of nearby bulk material, and induce additional energy dissipation via mechanisms such as viscoelasticity. In the current work, our model PSA is fabricated using a nearly hysteresis-free elastomeric bulk, such that we remove most of the bulk energy dissipation and focus on the measured interfacial fatigue threshold and slow crack threshold. More discussions about the consequence of this hysteresis-free bulk will be provided in the later part of the paper.

**Figure 4a** illustrates the near-zero hysteresis of the bulk PBA without dip coating under 50 cycles of uniaxial stress-stretch tests. The hysteresis area (between the loading and unloading curves) in the first cycle is 1.67 % of the total area under the loading curve, indicating minimum bulk dissipation. Despite the near-zero bulk hysteresis of the PBA elastomer, its adhesion toughness to PET significantly increases from about 11 $J/m^2$ to about 400 $J/m^2$ after dip coating with the viscoelastic sticky layer (**Figure 4b**). More detailed material characterizations and discussions have been reported in the previous work [38].

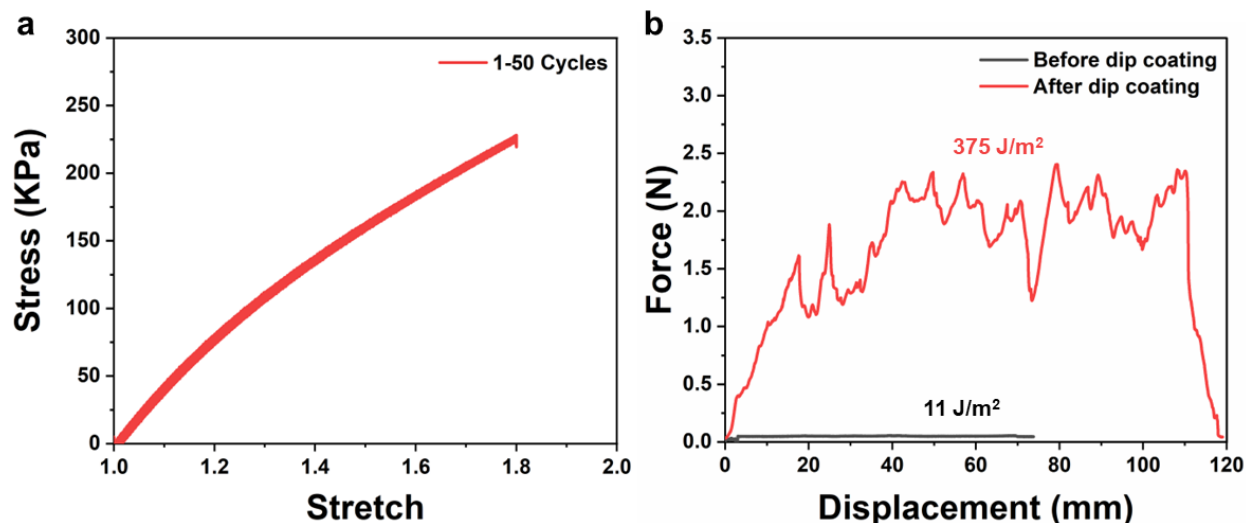

Figure 4. Mechanical characterizations of bulk elastomer and surface layer. (a) Cyclic stress-stretch curve of the bulk PBA shows negligible hysteresis. (b) Peeling tests of the PBA elastomer with PET adherend before and after dip coating.

**4. Comparison between cyclic and monotonic peeling**

In cyclic peeling, the tensile tester prescribes a constant peak peeling force in each cycle, and the peeling force is converted to the energy release rate. **Figure 5a** illustrates the experimental raw data of the measured peak peeling force over thousands of cycles, showing a relatively good agreement between the loading machine and the programmed force-controlled loading profile. Under each prescribed peeling force or energy release rate $G$, the crack length is directly recorded by the tensile tester due to the inextensible backing layer (**Figure 5b**). In all tests, we observe an initial transient state with nonlinear crack growth in the first few thousands of cycles. After the transient state, a steady state emerges, where the slope is nearly constant. As the crack further grows and approaches the end of the bilayer, nonlinear crack growth is observed again, possibly due to the deviation from the steady-state peeling geometry (**Figures 2a&3a**). We will use the constant slope in the steady state as the crack growth per cycle for the following analysis.

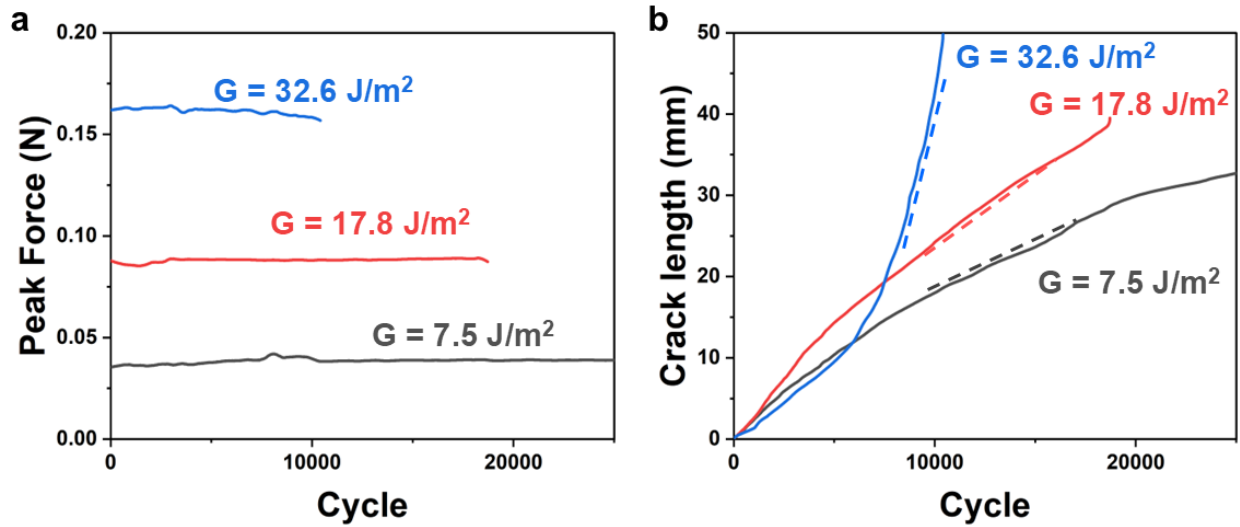

Figure 5. Experimental raw data of (a) the peak force and (b) the crack length over peeling cycles in the force-controlled cyclic peeling tests. The corresponding energy release rate for each curve is also listed. The dashed lines represent the steady-state crack growth with nearly constant slopes.

In monotonic peeling, the crack speed is prescribed as a constant by the tensile tester. The recorded force-displacement curves under different crack speeds are illustrated in **Figure 6**. We use the plateau of each curve to calculate the steady-state energy release rate $G$. As expected, the energy release rate increases with the crack speed, indicating the rate-dependent debonding of PSA, possibly due to the viscoelasticity of its sticky surface layer.

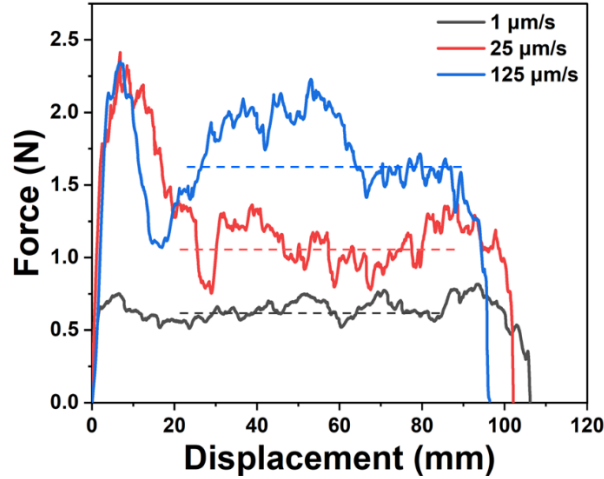

Figure 6. Force-displacement curves at different crack speeds in monotonic peeling tests. The dashed lines represent the steady-state crack growth plateaus.

We next plot the crack growth per cycle d$c$/d$N$ as a function of the energy release rate $G$ in cyclic peeling (**Figure 7a**), and plot the crack speed $v$ against $G$ in monotonic peeling (**Figure 7b**). In the two plots, we use linear extrapolation to obtain the interfacial fatigue threshold $\Gamma_0$ and the slow crack threshold $G_0$, respectively.

For the same bilayer formed by the model PSA and PET, the interfacial fatigue threshold is measured to be $\Gamma_0$ = 4.6 J/m$^2$, much lower than the slow crack threshold $G_0$ = 33.9 J/m$^2$, both further below the adhesion toughness around 400 J/m$^2$ measured at a finite crack speed. As expected, PSAs suffer from fatigue fracture, just like many other soft materials and adhesion systems [17, 22]. Moreover, our experimental data show that the slow crack threshold $G_0$ = 33.9 J/m$^2$ is much larger than the interfacial fatigue threshold $\Gamma_0$ = 4.6 J/m$^2$ by one order of magnitude, even though the hysteresis-free bulk PSA has already removed most of the bulk dissipation. Motivated by these observations, we will next develop a modified Lake-Thomas model for the measured $\Gamma_0$ resulting from noncovalent interfacial bond breaking, and then provide further discussions on the discrepancy between $\Gamma_0$ and $G_0$.

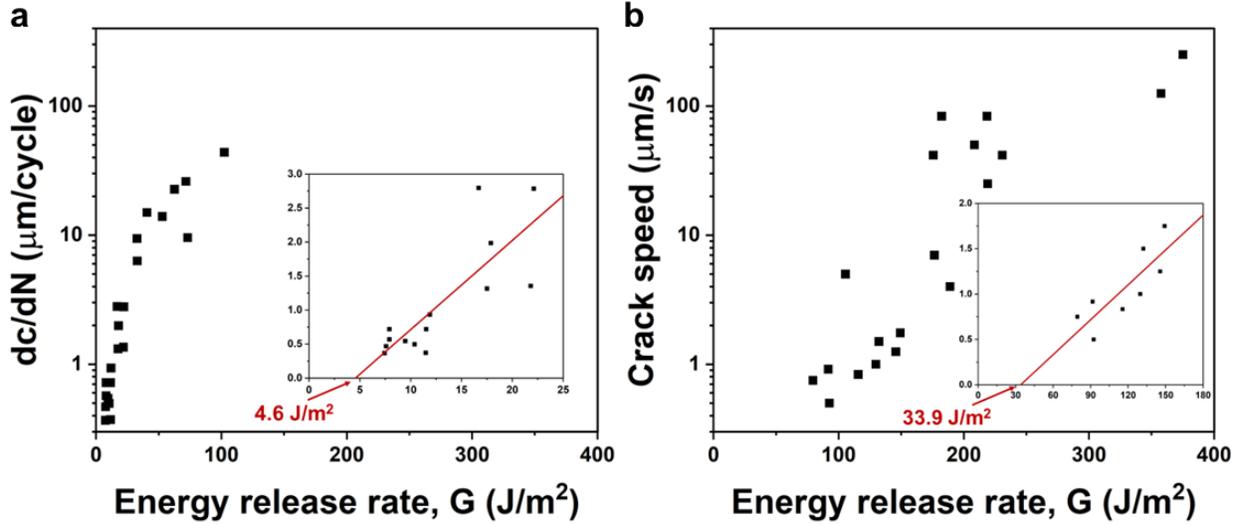

Figure 7. Experimental data on (a) crack growth per cycle d$c$/d$N$ as a function of the energy release rate $G$ in cyclic peeling, and (b) crack speed $v$ as a function of energy release rate $G$ in monotonic peeling. The interfacial fatigue threshold $\Gamma_0$ and the slow crack threshold $G_0$ are obtained using linear extrapolation of the data points shown in the inset.

## 5. Modified Lake-Thomas model for interfacial fatigue threshold

We develop a modified Lake-Thomas model to describe the interfacial fatigue threshold $\Gamma_0$ of a PSA. Our model builds on the classical Lake-Thomas model for the fatigue threshold of covalently crosslinked elastomeric networks [23], as well as a preliminary adaptation of the model for noncovalent networks in our previous paper [17].

We assume the following physical process of polymer chain debonding at the interfacial fatigue threshold (**Figure 8a**). During the formation of adhesion, most part of a surface polymer chain is sufficiently adhered to the adherend, forming many noncovalent bonds and bridging an interfacial crack. Debonding of one chain releases both the noncovalent bond energy and the entropic elastic energy of the chain. This slightly differs from the classical model for covalent networks, where Lake and Thomas assume that during fracture, every Kuhn segment carries a bond force equivalent to the covalent bond strength, and the bond energy in the entire chain is released. Based on this modified physical picture, we express the interfacial fatigue threshold as $\Gamma_0 = \Sigma(W_1 + W_2)$, where $\Sigma = \left(b^2\sqrt{n}\right)^{-1}$ is the number of surface chains per unit area [3, 23], $b$ is the Kuhn length, $n$ is the total Kuhn segments in a polymer chain, $W_1$ is the total noncovalent bond energy that bridges a chain at the interface, and $W_2$ is the elastic energy of a chain at debonding.

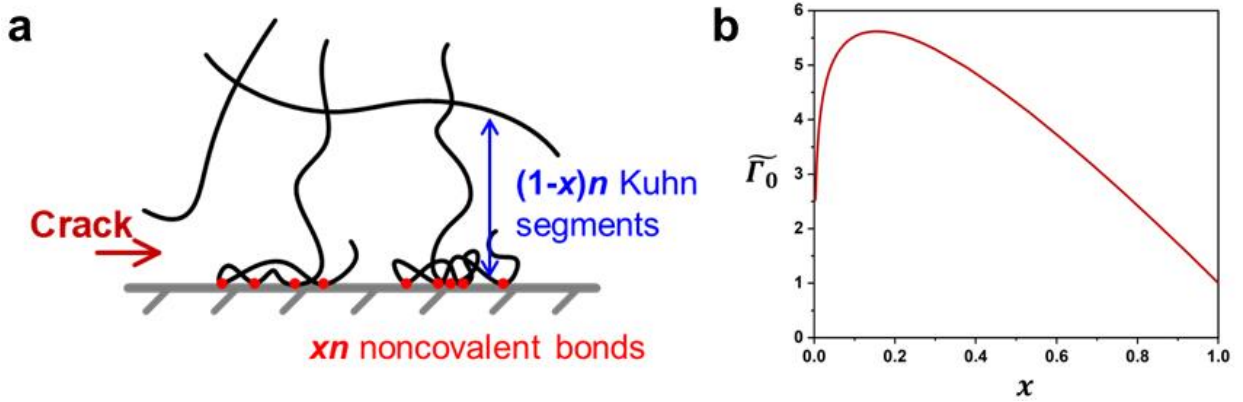

Figure 8. (a) Schematic of the modified Lake-Thomas model. (b) The calculated dimensionless function $\widetilde{\Gamma}_0(x)$.

We further assume that for each polymer chain attached to the adherend surface, there are $xn$ Kuhn segments noncovalently bonded at the interface ($0 \leq x \leq 1$), and $(1-x)n$ Kuhn segments left in the chain (**Figure 8a**). At debonding, the $xn$ noncovalent bonds are assumed to simultaneously break, releasing all their bond energies $W_1$ as well as the elastic energy $W_2$ in the polymer chain formed by the $(1-x)n$ Kuhn segments. Taking $f$ as the bond strength of one noncovalent bond, we express the total noncovalent bond energy $W_1$ as:

$$W_1 = xnfb, \qquad (1)$$

Using the freely jointed chain model [44, 45], the elastic energy of a chain at debonding is expressed as:

$$W_2 = (1-x)nkT\left[\frac{xn\beta}{\tanh(xn\beta)} + \log\left(\frac{xn\beta}{\sinh(xn\beta)}\right)\right], \qquad (2)$$

where $kT$ is the temperature in the unit of energy and $\beta = fb/kT$.

Finally, the interfacial fatigue threshold is expressed as

$$\Gamma_0 = \frac{\sqrt{n}kT}{b^2}\tilde{\Gamma}_0(n,x), \qquad (3)$$

with a dimensionless function

$$\tilde{\Gamma}_0(n,x) = x\beta + (1-x)\left[\frac{xn\beta}{\tanh(xn\beta)} + \log\left(\frac{xn\beta}{\sinh(xn\beta)}\right)\right]. \qquad (4)$$

For the current PSA, the weight ratio between the monomer BA (molecular weight $M_{BA} = 128.17$ g/mol and density $\rho_{BA} = 8.9 \times 10^5$ g/m$^3$) and monomer IBA (molecular weight $M_{IBA} = 208.3$ g/mol and density $\rho_{IBA} = 9.86 \times 10^5$ g/m$^3$) is 8:2. We also assume that a Kuhn segment has the same size as a monomer. Therefore, the average volume of a Kuhn segment is calculated as $V = (0.8/\rho_{BA} + 0.2/\rho_{IBA})/[(0.8/M_{BA} + 0.2/M_{IBA})N_A] = 2.54 \times 10^{-28}$ m$^3$. The average Kuhn length is thus $b = V^{\frac{1}{3}} = 0.633$ nm. Since the weight ratio of the crosslinker PEGDA ($M_{PEGDA} = 575$ g/mol) is 0.1:100 of the total weight of monomers, the average number of Kuhn segments between two crosslinkers is calculated as $n = (0.8/M_{BA} + 0.2/M_{IBA})/(10^{-3}/M_{PEGDA})/2 = 2070$. Finally, we assume that $\beta = fb/kT = 1$, representing the bond energy of a single noncovalent bond, such as the van der Waals bond.

With these, all the physical parameters in Eqs. (1)-(4) are obtained from experiments except for the unknown $x$. With $n = 2070$ fixed in the current study, the dimensionless $\tilde{\Gamma}_0(n,x)$ is reduced to $\tilde{\Gamma}_0(x)$. We observe that $\tilde{\Gamma}_0(x)$ is a nonmonotonic function of $x$, due to combination of the energetic bonding energy that scales with $x$ and the entropic elastic energy that scales with $(1-x)$. Therefore, we take the maximum of $\tilde{\Gamma}_0(x)$ as the input for Eq. (3), analogous to a strength-based failure criterion. As shown in **Figure 8b**, using the current values of parameters, $\tilde{\Gamma}_0(x)$ has a maximum of 5.62. Substituting all the parameters into Eq. (3), we obtain $\Gamma_0 = 2.6$ J/m$^2$. This theoretical estimation agrees well with the experimentally measured threshold $\Gamma_0 = 4.6$ J/m$^2$.

## 6. Discussion and summary

The huge discrepancy between the interfacial fatigue threshold $\Gamma_0$ (~ 3 J/m$^2$) and the finite-speed adhesion toughness $\Gamma$ (~ 400 J/m$^2$) stems from the multiscale energy dissipation during the interfacial fracture of PSAs. A classical equation in soft fracture and adhesion reflects this multiscale energy dissipation [3, 46-48], $\Gamma = \Gamma_0 [1 + f_d(v, T)]$, where $f_d(v, T) \geq 0$ is an amplifying factor accounting for the additional toughening beyond the crack processing zone. In literature, the fatigue threshold $\Gamma_0$ is also denoted as the *intrinsic toughness*, representing the minimum energy needed to propagate a crack. Most soft materials and adhesives have $\Gamma_0$ much smaller than $\Gamma$ by orders of magnitude. As a result, a systematic comparison between $\Gamma_0$ and $\Gamma$ has offered a useful lens to study the fundamental molecular processes and their consequences in fracture. Recent successful examples in other elastomers and gels include the identification of the bond dissociation origin [31, 49, 50], and quantification of the size of crack processing zone [41, 51]. In PSAs, by contrast, their interfacial fatigue fracture and the comparison between $\Gamma_0$ and $\Gamma$ have never been studied. This knowledge gap is likely due to their conventional application domain such as duct tape, which requires only static loads. As new PSAs expand their application domains in fields like wound dressing, wearable sensors, and flexible displays, research on their fatigue resistance under cyclic loads is expected to grow rapidly.

Our experiments systematically measure the interfacial fatigue threshold $\Gamma_0$, the slow crack threshold $G_0$, and the adhesion toughness $\Gamma$ of the same PSA-adherend bilayer. We find that even at the thresholds, $\Gamma_0 = 4.6$ J/m$^2$ is much lower than $G_0 = 33.9$ J/m$^2$. Substituting these values into the equation $\Gamma = \Gamma_0 [1 + f_d(v, T)]$, we conclude that the larger $G_0 = \Gamma(v = 0) = \Gamma_0 [1 + f_d(0, T)]$ indicates additional rate-independent energy dissipation, leading to $f_d(0, T) > 0$. Such a large difference between $\Gamma_0$ and $G_0$ has been previously observed in fracture of tough, double-network polyacrylamide/calcium-alginate hydrogels [41]. In those gels, we concluded that the solid-like calcium-alginate network can provide significant toughening even at the slow crack threshold where the crack speed approaches zero. However, different from those gels, the current model PSA has its bulk material nearly free of hysteresis. As a result, the additional dissipation in $G_0$ can only come from the sticky, highly viscoelastic surface layer. In a preliminary attempt to explain this additional dissipation, we relate it to the long-reported, notable fibril formation during the debonding of PSAs [4, 19, 52]. In experiments, we take photos at the crack tip in monotonic peeling with a crack speed of 2 μm/s (**Figure 9**), close to the slow crack threshold. We observe multiple small fibrils emerge and bridge the crack tip. Previous studies have shown that such fibrillar structures can effectively toughen the adhesion of a PSA at finite crack speeds due to nonlinear viscoelastic dissipation [19], originating from physical processes such as molecular friction [53], chain disentanglement [54], and chain pullout [55]. Nevertheless, the detailed toughening effect from fibrils at the zero-speed, slow crack threshold requires further investigation. In modeling, finite element simulation of monotonic peeling using a cohesive zone model [56] can incorporate different dissipations in the elastic bulk and viscoelastic surface layer, capture the behavior of interfacial debonding and fibril rupture, and explain the large discrepancy between $\Gamma_0$ and $G_0$.

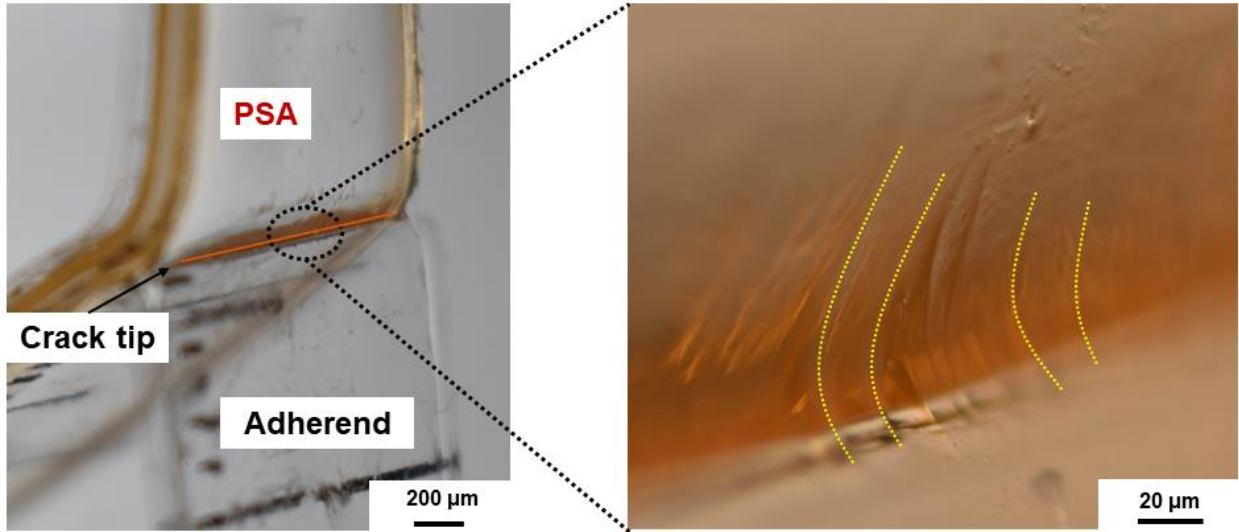

Figure 9. Photo of the crack tip in monotonic peeling with a crack speed of 2 μm/s, showing multiple small fibrils. Yellow dash lines on the right represent the structure of some fibrils.

In our theory, we assume that the $xn$ noncovalent bonds break simultaneously, and we take the maximum of $\tilde{\Gamma}_0(x)$ for the theoretical $\Gamma_0$. One possible physical picture associated with this assumption is described as follows. Before the crack propagation, most part of a surface polymer chain is bonded to the adherend, leading to a value of $x$ larger than the critical $x$ that maximizes $\tilde{\Gamma}_0$. As the local energy release rate increases, some noncovalent bonds break and reduce the value of $x$. This process is stable until $x$ reaches the critical $x$. Afterwards, further decreasing $x$ reduces $\tilde{\Gamma}_0$, leading to an unstable cascade of bond breaking towards complete chain debonding.

More experiments can be conducted to quantitatively validate the modified Lake-Thomas model. For example, $\Gamma_0$ of the same bilayer can be measured at different temperatures. The average number of Kuhn segments between two crosslinkers $n$ can be readily controlled by the concentration of the crosslinker PEGDA. The better comparison between the experimental and theoretical $\Gamma_0$ will provide valuable information for future improvement of the model.

In summary, we have conducted systematic experiments on cyclic peeling and monotonic peeling of a model PSA. The interfacial fatigue threshold and slow crack threshold are measured from the two tests respectively. The model PSA consists of a hysteresis-free elastomeric bulk dip-coated with a viscoelastic sticky surface layer. We find that the measured interfacial fatigue threshold ($\Gamma_0 = 4.6$ J/m$^2$) is much lower than the slow crack threshold ($G_0 = 33.9$ J/m$^2$), both further below the adhesion toughness (~ 400 J/m$^2$ at a finite crack speed). A modified Lake-Thomas model is derived to well describe the measured interfacial fatigue threshold. The large discrepancy between the two thresholds indicates rate-independent dissipation beyond the intrinsic toughness, possibly due to the sticky surface layer and the observed fibril structures emerging from it. It is hoped that this study will provide new fundamental knowledge for fracture mechanics of PSAs, as well as guidance for future tough and durable PSAs.

**Author contributions**
Y.W. and R.B. designed the research. Y.W. performed the research and analyzed the data. Y.W. and R.B. wrote the first draft of the paper. All authors contributed to the writing of the final manuscript.

**Conflict of interest**
The authors declare no conflict of interest.

**Acknowledgments**
This work was supported by the ASME (American Society of Mechanical Engineers) Haythornthwaite Research Initiation Grant and the ACS (American Chemical Society) Petroleum Research Fund.

**Movies**
Movie 1. Cyclic peeling of PSAs. A bilayer (6 cm ×1 cm ×1.5 mm) with a precut crack of 0.5 cm is stretched under a prescribed energy release rate between 0 and 100 J/m$^2$, with a crack speed of 250 μm/s.

Movie 2. Monotonic peeling of PSAs. A bilayer (6 cm ×1 cm ×1.5 mm) with a precut crack of 1 cm is monotonically peeled at a constant crack speed of 50 μm/s.